\documentstyle[12pt,epsfig,amssymb]{article}
\def\ap#1#2#3{Ann.\ Phys.\ (NY) #1 (19#3) #2}
\def\cmp#1#2#3{Commun.\ Math.\ Phys.\ #1 (19#3) #2}

\def\np#1#2#3{Nucl.\ Phys.\ B#1 (19#3) #2}

\def\pl#1#2#3{Phys.\ Lett.\ #1B (19#3) #2}
\def\pr#1#2#3{Phys.\ Rev.\ D #1 (19#3) #2}
\def\prb#1#2#3{Phys.\ Rev.\ B #1 (19#3) #2}
\def\prep#1#2#3{Phys.\ Rep.\ #1 (19#3) #2}

\def\lnc#1#2#3{Lettere al Nuovo Cimento #1 (19#3) #2}
\def\jmp#1#2#3{J.\ Math.\ Phys.\ #1 (19#3) #2}
\def\nc#1#2#3{Il Nuovo Cimento #1A (19#3) #2}

\def\frac#1#2{ {{#1} \over {#2} }}
\def\ie{\hbox{\it i.e.}{ }}

\def\half{\mbox{\small $\frac{1}{2}$}}
\def\ds#1{\ooalign{$\hfil/\hfil$\crcr$#1$}}
\def\gm{\gamma_{\mu}}
\def\gn{\gamma_{\nu}}
\def\gr{\gamma_{\rho}}
\def\gs{\gamma_{\sigma}}

\def\gc{\gamma_5}
\def\e#1#2#3#4{\varepsilon^{#1 #2 #3 #4}}
\def\re#1{(\ref{#1})}
\def\beq{\begin{equation}}
\def\eeq{\end{equation}}
\def\beeq{\begin{eqnarray}}
\def\beeqn{\begin{eqnarray*}}
\def\eeeq{\end{eqnarray}}
\def\eeeqn{\end{eqnarray*}}
\def\m{\mu}
\def\n{\nu}
\def\r{\rho}
\def\s{\sigma}
\def\a{\alpha}
\def\b{\beta}

\def\G{\Gamma}
\def\eps{\epsilon}
\def\L{\Lambda}
\def\l{\lambda}
\def\g{\gamma}
\def\D{\Delta}
\def\bc{\bar c}
\def\bchi{\bar \chi}

\def\bpsi{\bar \psi}
\def\n{\nu}
\def\m{\mu}
\def\r{\rho}
\def\s{\sigma}

\def\G{\Gamma}
\def\bG{\bar\Gamma}
\def\eps{\epsilon}
\def\L{\Lambda}
\def\l{\lambda}
\def\bl{\bar\lambda}
\def\g{\gamma}
\def\de{\delta}
\def\D{\Delta}
\def\t,#1{t^{#1}}
\def\f,#1#2#3{f^{#1 #2 #3}} 
\def\tr{\,\mbox{Tr}\,}
\def\se{S_{\mbox{\footnotesize{eff}}}}
\def\st{S_{\mbox{\footnotesize{tot}}}}

\def\si{S_{\mbox{\footnotesize{int}}}}
\def\sbrs{S_{\mbox{\footnotesize{BRS}}}}

\def\Pir{\Pi_{\mbox{\footnotesize{rel}}}}

\def\Piinv{\Pi_{\mbox{\footnotesize{inv}}}}
\def\Pit{\tilde\Pi}

\def\Gir{\G_{\mbox{\footnotesize{irr}}}}
\def\De{\D_{\mbox{\footnotesize{eff}}}}

\def\DG{\D_{\G}}
\def\DGr{\D_{\G,\mbox{\footnotesize{rel}}}}
\def\DGi{\D_{\G,\mbox{\footnotesize{irr}}}}
\def\DGb{\bar{\D}_{\G}}
\def\DGh{\hat{\D}_{\G}}
\def\DGhr{{\hat{\D}}_{\G,\mbox{\footnotesize{rel}}}}

\def\LdL{\L\partial_\L}
\def\UV{$\L_0\to\infty\;$}

\def\bit{\begin{itemize}}
\def\eit{\end{itemize}}
\def\ben{\begin{enumerate}}
\def\een{\end{enumerate}}

\def\nome#1{{\label{#1}}}
\def\p{\partial}
\def\K{K_{\L\L_0}}

\def\Kiu{K_{0\L_{0}}}
\def\Kin{K_{\L_{0}\infty}}
\def\bom#1{\mbox{\boldmath$#1$}}

\begin{document}
\begin{titlepage}

\begin{flushright}
UPRF-97-07 \\
\end{flushright}
\vspace{.4in}
\begin{center}
{\large{\bf Chiral gauge theories and anomalies

in the Wilson renormalization group approach}}
\bigskip \\ M. Bonini and F. Vian
\\
\vspace{\baselineskip}
{\small Universit\`a degli Studi di Parma \\ and\\
I.N.F.N., Gruppo collegato di Parma, 
\\viale delle Scienze, 
43100 Parma, Italy} \\
\mbox{} \\
\vspace{.5in}
{\bf Abstract} 
\bigskip 
\end{center} 
\setcounter{page}{0}
\begin{quote}
We extend the Wilson renormalization group (RG) formulation to chiral gauge
theories and show that local gauge symmetry can be implemented
by a suitable choice of the RG flow boundary conditions.
Since the space-time dimension is four, there is no ambiguity 
in handling the matrix $\g_5$ and left and right fermions are not 
coupled.  As a result the ultraviolet action contains all possible 
globally chiral invariant interactions.
Nevertheless, the correct chiral anomaly is reproduced.

PACS number(s): 11.10.Gh, 11.15.-q, 11.30.Rd
\end{quote}

\end{titlepage}

\section{Introduction}
The problem of finding a consistent renormalization procedure of a 
chiral gauge theory (CGT) is still an active field of investigation,
both in the framework of perturbation theory \cite{chiral} and in
lattice regularization \cite{roma}. In the presence of chiral fermions
no regularization is known to preserve  chiral symmetry.
This is not  a technical problem but it is related to the fact that
chiral symmetry is anomalous \cite{abj}. In dimensional regularization with 
minimal subtraction \cite{eps} the
correct definition of the matrix $\g_5$ produces
chiral breaking terms \cite{gamma5}, although they formally vanish  for 
$d\to4$.
The appearance of these breaking terms is necessary to reproduce the
anomaly for external currents.  In order to ensure the
Slavnov-Taylor (ST) identites of the local chiral symmetry for the
renormalized theory, the minimal subtraction procedure must be
supplemented by additional finite non-invariant counterterms 
\cite{chiral,RSS,F}.
Similarly, in
lattice regularization one must introduce the Wilson term in
order to avoid the fermion doubling and reproduce the correct
anomaly.  This term explicitly breaks chiral invariance and one
adds all possible counterterms to the naive action to compensate
the explicit symmetry breaking induced by the Wilson term \cite{roma}.
Notice that in both approaches the regularized Lagrangian
couples left and right fermions and therefore also global chiral
symmetry is broken.

The proof that it is possible to compensate the breaking due to
the regularization by a suitable choice of non-invariant counterterms
in the Lagrangian is an essential ingredient of the renormalization
procedure. If the fermionic content of the theory ensures the
cancellation of the one-loop chiral anomaly, this fine-tuning problem
has a solution.
Its solvability is independent of the regularization procedure since the 
classification of all possible anomalies is an algebraic problem
\cite{cohom, Becchi0}.

Recently it has been shown that the Wilson renormalization group (RG) 
formulation \cite{W,P} can be used to deduce
the ST identities in the Yang-Mills theory. In this method one
introduces an ultraviolet (UV) and infrared (IR) cutoff, $\L$ and 
$\L_0$ respectively, in the propagator and studies the evolution 
in $\L$ of the cutoff effective action.  
In this way  gauge invariance is explicitly broken by the
cutoff propagator and  one proves that, by properly fixing the boundary
conditions of the RG equation, the ST identities can be satisfied when 
all cutoffs
are removed (at least in perturbation theory).  This has been shown
for the pure YM case both in terms of  the ``bare'' couplings of the effective
action at the ultraviolet scale \cite{B} and of the physical couplings
\cite{noi}. 
In this  case the possibility of solving the fine-tuning problem 
is guaranteed by the fact that, once the renormalization conditions 
are imposed and all cutoffs are removed, the renormalized Green 
functions are independent of the regularization procedure 
and, in particular, they coincide with those obtained via 
dimensional regularization.

The RG formulation con be trivially extended to chiral gauge
theories.  Since in this case there is no regularization which
preserves the symmetry, the RG formulation has no disadvantages
with respect to other procedures. As a matter of fact, in the RG
approach the space-time dimension is $d=4$ so that there is no
ambiguity in the definition of the matrix $\g_5$ and in the
regularized action left and right fermions are not coupled.
Therefore the solution of the fine-tuning procedure is
simpler than in the standard cases (\ie dimensional or lattice
regularization), since the possible counterterms must be
globally chiral invariant.  Although there is no left-right
coupling, we will show that one obtains the correct chiral
anomaly (if the matching conditions for the anomaly cancellation
are not fulfilled).

The paper is organized as follows. In sect. 2 we set up the RG formalism
for the CGT and discuss the boundary conditions for the flow
equations. In sec. 3 we schematically rederive the effective ST identities
for this theory and  solve the fine-tuning equation at
one-loop order and at the UV scale. Technical details of this
calculation are given in appendix A.
In sect. 4 we compute the one-loop chiral anomaly. 
Sect. 5 contains some conclusions.

\section{Renormalization group flow and effective action}
Consider the $SU(N)$ chiral gauge theory described by the classical 
Lagrangian (in the Feynman gauge)
\beq\nome{Scl}
S_{cl}= S_{YM} + S_F
\eeq
\beeq\nome{YM}&&
S_{YM}=\int d^4x
\,
\left\{
-\frac 1 4 F^a_{\mu\nu}  \, F^{a\mu\nu}
- \frac 1 2 \left(\p^\mu A^a_\mu \right)^2
- \bc^a \p^\mu D_\mu^{ab} c^b
\right\}\,,
\nonumber
\\ 
&&S_F=\int d^4x
\left(\bpsi_{L}\,i\, \ds{D}\, \psi_{L} + \bpsi_{R}\,i\, \ds{\p}\,
\psi_{R}\right)
\eeeq
where the gauge stress tensor and the covariant derivatives are given by
$F^a_{\mu\nu}=\p_\mu A^a_\nu -\p_\nu A^a_\mu
+ g \, f^{abc} A^b_\mu \,A^c_\nu$, 
$D^{ab}_\mu c^b =\p_\mu c^a + g \, f^{abc} A^b_\mu \, c^c$ and 
$D_\mu \psi =\p_\mu \psi + g \,A_\mu^a\, t^a \, \psi$, with $t^a$ 
the SU(N) generators in the appropriate representation, such that 
$[t^{a}, t^{b}]=f^{abc} t^{c}$ and 
$\tr [t^{a} t^{b}]=\half\de^{ab}$. 
This action is invariant under the BRS transformations \cite{BRS}
$$
\de A^a_\mu= \frac 1 g \eps \,D^{ab}_\mu c^b
\,,
\;\;\;\;\;\;\;
\de c^a= -\half\,\eps\, f^{abc} c^b \,c^c
\,,
\;\;\;\;\;\;\;
\de \bc^a= - \frac 1 g \,\eps\, \p^\mu A^a_\mu
\,,
$$
$$
\de \psi_{L} = - \eps\, c^a t^a\, \psi_{L}\,, 
\;\;\;\;\;\;\;
\de \bpsi_{L} = - \eps \,\bpsi_{L}\, c^a t^a\,, 
\;\;\;\;\;\;\;
\de \bpsi_{R} = \de \psi_{R} = 0
$$
with $\eps$ a Grassmann parameter.

Introducing the sources $u^a_\mu$, $v^a$, $\bl$ and $\l$  associated to 
the variations of $A^a_\mu$, $c^a$, $\psi$ and $\bpsi$ respectively 
one has the BRS action
\beq\nome{Stot}
S_{BRS}[\Phi_I,\g_i]
=
S_{cl} +
\int d^4x
\left\{
\frac 1 g u^a_\mu  D^{ab}_\mu c^b - \half f^{abc} v^a\,c^b \,c^c
+ \bl \,c^a\, t^a\, \psi_{L} +\bpsi_{L}\, c^a\, t^a\,\,  \l
\right\}
\eeq
where we have denoted by $\Phi_I$ and $\g_i$ the fields and the BRS 
sources
$$
\Phi_I=\{\,A^a_\mu, \,c^a, \,\bc^a, \,\bpsi,\,\psi\}
\,,
\;\;\;\;\;\;\;\;
\g_i=\{w_\mu^a,\, v^a, \,\bl, \, \l\}
\,,
$$
and  $w_\mu^a=u_\mu^a+g\partial_\mu \bc^a$ (no source is introduced for 
$\bc^a$).

In order to quantize the theory one
needs a regularization procedure of the ultraviolet divergences.
We regularize these divergences by assuming that in the path integral
one integrates only the fields with frequencies smaller than a given
UV cutoff $\L_0$. This procedure is equivalent to assume that the free
propagators vanish for $p^2 > \L_0^2$.

The generating functional is
\beq\nome{W}
Z[J,\g]=e^{iW[J,\g]}=\int {\cal D}\Phi \, \exp{i\{-\half (\Phi, 
D^{-1}\Phi)_{0\L_0}+(J,\Phi)_{0\L_0}+\si[\Phi,\g;\L_0]\}}
\,,
\eeq
where the free propagators are collected in the matrix $D^{-1}_{IJ}$ and,
more generally,
we introduce the cutoff scalar products between fields and sources
\beeq&&
\half (\Phi,\, D^{-1}\, \Phi)_{\L\L_0}\equiv
\int_p\, \, K^{-1}_{\L\L_0}(p)
\big\{p^2 \left[\half A^a_\mu(-p)\,A^a_\mu(p)
- \bc^a(-p)\, c^a(p)\right] 
\nonumber
\\
&&
\;\;\;\;\;\;\;\;\;\;\;\;\;\;\;
\;\;\;\;\;\;\;\;\;\;\;\;\;\;\;\;\;\;\;\;\;\;\;\;\;\;\;\;
-\bpsi_L(-p)\,\ds{p}\,\psi_L(p)
-\bpsi_R(-p)\,\ds{p}\,\psi_R(p)
\big\}\,,
\eeeq
\beeq
&&(J,\Phi)_{\L\L_0}\equiv
\int_p \, K^{-1}_{\L\L_0}(p) 
\big\{j^a_\mu(-p)\, A^a_\mu(p) +
[\bchi^a(-p) - \frac i g p_\mu u^a_\mu(-p)] \, c^a(p) +
\bc^a(-p)\, \chi^a(p) 
\nonumber 
\\
&&
\;\;\;\;\;\;\;\;\;\;\;\;\;\;\;\;\;\;\;\;\;\;\;\;\;\;\;\;
+\bpsi(-p) \eta(p) 
+\bar\eta(-p) \psi(p) 
\big\}
\,,
\eeeq
where $\int_p \equiv \int d^4p/(2\pi)^4$ and 
$K_{\L\L_0}(p)$ is a cutoff function which is one
for $\L^2 < p^2 < \L_0^2$ and rapidly vanishes outside.

The introduction of the cutoff function in the propagators breaks the 
gauge invariance, therefore the UV action $\si$ must contain all possible 
renormalizable interactions which are needed to compensate the symmetry 
breaking induced by the regularization. These are given by the monomials
in the fields, BRS sources and their derivatives which have dimension
not larger than four and are Lorentz and $SU(N)$ scalars, since Lorentz 
and global chiral gauge symmetries are  preserved by the cutoff 
regularization.
The independent monomials of the Yang-Mills sector are 
\beeq\nome{mon1}
&&
\tr(A_\mu \, A_\mu) \,,
\;\;\;\;\;\;\;
\tr(\partial_\nu A_\mu \, \partial_\nu A_\mu) \,,
\;\;\;\;\;\;\;
\tr(\partial_\mu A_\mu \, \partial_\nu A_\nu) \,,
\;\;\;\;\;\;\;
\tr(w_\mu \, \partial_\mu c )\,,\nonumber
\\
&&
\tr(A_\mu\, A_\mu \, \partial_\nu A_\nu) \,,
\;\;\;\;
\tr(A_\mu\, A_\nu \, \partial_\mu A_\nu) \,,
\;\;\;\;
\tr(w_\mu \, c \, A_\mu) \,,
\;\;\;
\tr(v \, c \, c )\,, 
\nonumber \\
&&
\tr(A_\mu \, A_\nu\, A_\mu \, A_\nu)\,,
\;\;\;\;\;\;
\tr(A_\mu \, A_\mu\, A_\nu \, A_\nu)\,,\nonumber
\\
&&
\tr(A_\mu \, A_\nu)\, \tr(A_\mu \, A_\nu)\,,
\;\;\;\;\;\;
\tr(A_\mu \, A_\mu)\, \tr(A_\nu \, A_\nu)\,,
\eeeq
where $A_\m=A^a_\m\, t^a$, $c=c^a\, t^a$, $w_\m=w^a_\m\, t^a$, 
$v=v^a\, t^a$ and the trace is over all possible representations. 
Notice that in the $SU(2)$ case there are  only one monomial with three
gauge fields and only two independent monomials with four gauge fields, while
for $SU(3)$ only three monomials with four gauge fields are independent. 
In the fermionic sector one has the five monomials 
\beq\nome{mon2}
\bpsi_{L}\, \ds{\p}\, \psi_{L}\, , 
\;\;\;\;\;\;
\bpsi_{R} \,\ds{\p}\, \psi_{R}\, ,
\;\;\;\;\;\;
\bpsi_{L}\, \ds{A^a}t^a\, \psi_{L} \, ,
\;\;\;\;\;\;
\bl \,c\, \psi_{L} \, ,
\;\;\;\;\;\;
\bpsi_{L}\, c\,  \l\,.
\eeq
All these interactions appear in $\si$ with couplings of 
non-negative dimension (relevant parameters) which depend on $\L_0$.
In order to obtain the physical theory one has to show that
these relevant parameters can be fixed in such a way that:
\newline
(1) the \UV limit can be taken by fixing the physical parameters such
as the masses, the coupling $g$ and the wave function normalization 
constant at a subtraction point $\mu$.  Perturbative renormalizability 
ensures that this can be done \cite{P,B,noi} (for this reason the 
dependence on the ultraviolet cutoff $\L_0$ will be sometimes understood).
\newline
(2) in the \UV limit the Slavnov-Taylor identities must be satisfied.
For a chiral gauge symmetry this requires that the gauge group is 
anomaly free or, more generally, constraints the fermionic content of 
the theory.
This is the crucial point we want to discuss in this paper.

According to Wilson one integrates over the fields with frequencies
$\L^2<p^2<\L_0^2$ and obtains
\beq\nome{Z'}
e^{iW[J,\g]}=N[J;\L,\L_0]\int {\cal D}\Phi \, 
\exp{i\biggl\{
-\half (\Phi, \,D^{-1}\Phi)_{0\L}+(J\,\Phi)_{0\L}
+\se[\Phi,\g;\L]
\biggr\}}
\,,
\eeq
where $N[J;\L,\L_0]$ contributes to the quadratic part of $W[J,\g]$
and $\se[\Phi,\g;\L]$ is the so called Wilsonian effective action.

This functional, apart from the tree level two point function,
is the generating functional of the connected 
amputated cutoff Green functions, in which the free propagators 
contain $\L$ as an infrared cutoff \cite{P, B}. 
In other words the functional 
\beq\nome{WL}
W[J',\g;\L]= 
\se[\Phi,\g;\L]+\half (\Phi,\, D^{-1}\Phi)_{\L\L_0} \,,
\;\;\;\;\;\;\;\;\;
J_I'(-p)=\K^{-1}(p)\,\Phi_J(-p)\, D^{-1}_{JI}(p) \,
\eeq
is the generator of the cutoff connected Green functions.
The so called ``cutoff effective action'' is given by 
the Legendre transform of $W[J,\g;\L]$
\beq\nome{Leg}
\G[\Phi,\g;\L]=W[J,\g;\L]-\int_p J\Phi\,.
\eeq
This functional generates the cutoff vertex functions in which the 
internal 
propagators have frequencies in the range $\L^2<p^2<\L_0$ and reduces to
the physical quantum effective action in the limits $\L\to 0$ and \UV
\cite{BDM, Wett}.
 
Taking into account that the variable $\L$ enters as a cutoff in
the internal propagators one derives the exact RG equation \cite{noi}
\beq\nome{eveq}
\LdL \Pi[\Phi,\g;\L]= -\frac{1}{2} 
\int_q [\LdL K_{\L\L_0}^{-1}(q)] 
D^{-1}_{IJ}(q)
\left(\frac1{\G_2(q;\L)} \cdot \bG[-q,q;\Phi,\g;\L]
\cdot \frac1{\G_2(q;\L)} \right)_{JI} \,,
\eeq
where the indices $I$ and $J$ run over the indices of the fields $\Phi$ and
\beq\nome{pigreco}
\Pi=\G+\half (\Phi,\,D^{-1}\Phi)_{\L\L_0}-
\half (\Phi,\,D^{-1}\Phi)_{0\L_0}
\eeq 
is the cutoff effective action in which 
the infrared cutoff in the free 
propagators has been removed. The auxiliary functional 
$\bG_{IJ}[q,q';\Phi,\g;\L]\,$ 
is given by the integral equation
\beeq \nome{bG}
&&
\bG_{IJ}[q,q';\Phi,\g;\L] = 
(-)^{\delta_J}\,\G_{IJ}^{int}[q,q';\Phi,\g;\L]\,\nonumber
\\
&&\;\;\;\;\;\;\;\;\;\;\;\;\;\;
-\int_{q''} \bG_{KJ}[-q'',q';\Phi,\g;\L]\,
 \left(\frac1{\G_{2}(q'';\L)}\right)_{LK}\,
\G^{int}_{IL}[q,q'';\Phi,\g;\L]
\,,
\eeeq
where $\delta_J$ is the fermionic number and the two point functions 
$\G_{2,\,IJ}$ and the functional $\G_{IJ}^{int}$ are defined by
$$
(2\pi)^8\frac {\delta^2\G[\Phi,\g;\L]} {\delta\Phi_J(q)\delta\Phi_I(q')}
=
(2\pi)^4\delta^4(q+q') \, \G_{2,\,IJ}(q;\L)+
\G_{IJ}^{int}[q',q;\Phi,\g;\L]\,.
$$
Equation \re{eveq} is non-perturbative but can be perturbatively solved 
once the boundary conditions are given at some $\L$.
We now discuss the crucial topic of the boundary conditions which 
provide the starting point for the loop expansion. 

\subsection{Boundary conditions: physical parameters and symmetry}
In order to set the
boundary conditions it is useful to distinguish
between relevant and irrelevant vertices.  The relevant part of the
cutoff effective action involves only monomials in the fields and
sources with dimension not larger than four
\beq\nome{pir}
\Pir[\Phi,\g;\s_i(\L)]=\sum_i \s_i(\L)\,P_i[\Phi,\g]\,,
\eeq
where the sum is over the monomials $P_i(\Phi,\g)$  given in 
\re{mon1} and \re{mon2}.
The couplings $\s_i(\L)$ can be expressed in terms of the 
cutoff vertices at a given subtraction point (see  \cite{noi} for
the procedure of
extracting the relevant part from a given functional with a
non-vanishing subtraction point in the pure gauge case and \cite{chi} for 
the fermionic sector).

The remaining part of the cutoff effective action 
is called ``irrelevant''. 
Since we expect the theory to be renormalizable, for $\L\sim\L_0$
the dimension of the irrelevant couplings should be given only by 
negative powers of $\L_0$. Thus the simplest boundary condition for the 
irrelevant part of the cutoff effective action is
\beq\nome{irr}
\Gir[\Phi,\g;\L=\L_0]=0\,.
\eeq
For $\L=\L_0$, then, the cutoff effective action becomes local and
corresponds to the bare action $\si$ in \re{W}, with the bare
couplings given by  $\s_i(\L_0)$. 
The  way in which the boundary conditions for the couplings $\s_i(\L)$ 
are determined is not straightforward. 

In a previous paper we set the boundary conditions for these couplings
at the physical point $\L=0$. In this way some of the relevant 
couplings are related to physical couplings such as
the wave function normalizations and the three-vector coupling $g$ at
a subtraction point $\mu$. The remaining are fixed imposing the
symmetry at the physical point. This procedure is highly not trivial
since one has to analyse  non-local functionals. 

Alternatively one can discuss the symmetry at the ultraviolet scale
and determine $\s_i(\L=\L_0)$.  In this case the discussion is
simpler, since all functionals are relevant, but one has to perform a
perturbative calculation (\ie to solve the RG equations)  to obtain
the physical couplings. Notice that while the physical couplings 
are independent of the cutoff function, the bare action is generally not. 

In this paper we consider this second possibility,
although  the wave function normalizations and the gauge coupling $g$
at a subtraction point $\mu$ are still set at $\L=0$. As a matter of
fact  there are combinations of the monomials in \re{pir} which 
are not involved in the fine-tuning, so that the corresponding 
couplings are free and can be fixed at the physical point $\L=0$.
Before explaining the details of the fine-tuning procedure we recall how 
to implement the gauge symmetry in the RG formulation.

\section{Effective ST identities} 
The gauge symmetry requires that the effective action satisfies the ST 
identities ${\cal S}_{\G'} \G'[\Phi,\g]=0$, where 
$\G'[\Phi,\g]=\G[\Phi,\g]-\int d^4x \, \frac 12 (\p^\m A^a_\m)^2$ and 
$$
{\cal S}_{\G'}=  \int_p \left( \frac{\de \, \G'}{\de \Phi_i(-p)}
\frac{\de\, }{\de \g_i(p)} + \frac{\de \, \G'}{\de \g_i(p)} 
\frac{\de\, }{\de \Phi_i(-p)}\right)
$$
is the Slavnov operator \cite{Becchi0}. The ST
identities can be directly formulated for the Wilson effective action
$\se$ at any $\L$. 
In the generating functional \re{Z'} one performs  \cite{B,noi,Ellw,MT}
the following cutoff change of variables~\footnote{
Here and in the following  we use the notation $\Phi_i$ when the set of 
fields $\Phi$ does not include $\bc$.}
$$
\de\Phi_i(p)=\eps K_{0\L}(p) \frac{\de\st}{\de\g_i(-p)}\,,
\;\;\;\;\;\;\; \de\bc(p)=\eps\frac i g p_\mu A_\mu \,,
$$ where $\eps$ is a Grassmann parameter and $\st$ is the total action
(\ie $\se$ plus the source and the quadratic terms in \re{Z'}). Then
one gets the identity
\beeq \nome{sj}
&& \exp{iW[J,\g]}\,{\cal S}_J \cdot W[J,\g] =N[J;\L,\L_0]\nonumber
\\
&& \times
\int{\cal D}\Phi\;\De[\Phi, \g; \L, \L_0] \; 
\exp i{\biggl\{
-\half (\Phi, \,D^{-1}\Phi)_{0\L}
+\se[\Phi,\g;\L,\L_0]
+(J,\,\Phi)_{0\L} 
\biggr\}}
\eeeq
${\cal S}_J$ being the usual Slavnov operator \cite{Becchi0}.
Therefore in the RG formulation  the gauge symmetry condition is $\De=0$
(for the explicit form of $\De$, see for instance \cite{noi}).

In perturbation  theory one can show that  $\De$ satisfies a linear 
evolution equation, so that $\De=0$ at any $\L$ provided it vanishes 
at some $\L_R$ \cite{B,noi}. 
This is ensured by fixing the couplings in $\se$ at
the scale $\L_R$.  The solution of this fine-tuning problem needs the
so-called consistency conditions, which are a set of algebraic
identities coming from the nihilpotency of ${\cal S}_J$
(see ref.~\cite{B} for a detailed analysis of the pure gauge SU(2) 
model in the case $\L_R\ne0$ and \cite{noi} in the case $\L_R=0$). 
 
Such analysis is simplified if one removes all reducible 
contributions from $\De$ and studies the operator $\DG$ 
\cite{Ellw,MT} obtained from $\De$ through a Legendre transform. 
It is convenient to write $\DG$ as a sum of two contributions 
$\DG=\DGb+\DGh$ which are defined as follows 
\beq\nome{dgb}
\DGb[\Phi,\g;\L]=-\int d^4 p \; \Kiu (p)\,
\frac{\de\Pi'[\Phi,\g;\L]}{\de\Phi_i(-p)}\frac{\de\Pi'[\Phi,\g;\L]}{\de\g_i(p)}
\eeq
and 
\beeq\nome{dgh}
&&\DGh[\Phi,\g;\L]
=i\hbar\,
\int_{p,q} K_{0\L}(p)\left\{(-1)^{\de_L} \left(
\frac1{\G_2(q;\L)}\,\bG[-q,-p;\L]\right)_{JL} -\,\de_{JL}\, 
\de^4(p+q)\right\}
\nonumber\\&&
\;\;\;\;\;\;\;\;\;
\times 
\left(\frac1{\G_2(-p;\L)D_{\L \L_0} (-p)}\right)_{Li} \, 
\, \frac{\de^2}{\de\Phi_J(q)\de\g_i(p)}
\left(\Pi[\Phi,\g;\L]-\frac1g \int_{x}u_{\m}\p_{\m}c\right)
\,,
\eeeq
where $\bG$ is given in \re{bG} and $\Pi'$ is the expression obtained
by removing  the gauge fixing term in \re{pigreco}.
In  \re{dgh} we have inserted the factor $\hbar$ to put into evidence 
that this term vanishes at tree level.
Notice that at $\L=0$ and in the UV limit the gauge symmetry condition $\DG=0$
reduces to the ST identities, since $\Pi'$ becomes $\G'$ and $\DGh$ vanishes.

One can show \cite{MT} that 
the evolution of the vertices of $\DG$ at loop ${\ell}$ depends on
vertices of $\DG$ itself at lower loop order, so that if 
$\DG^{(\ell')}=0$ at any loop order $\ell'<\ell$, then
\beq\nome{chie}
\LdL\DG^{(\ell)}=0\,.
\eeq
Thus one can choose to analyse $\DG$ at any $\L$, in particular at $\L=\L_R$ 
much bigger than the subtraction scale $\mu$. In the following $\L_R$ will be 
identified with the UV scale $\L_0$. The functional $\DGh(\L_0)$ is 
schematically represented in figure~1. Notice that in \re{dgh}
the term 
proportional to $\de_{JL}$ does not contribute since in $\Pi[\Phi,\g;\L_0]$ 
diagonal interactions between a field and its own source are absent.

\begin{figure}[htbp]
\epsfysize=4cm
\begin{center}
\epsfbox{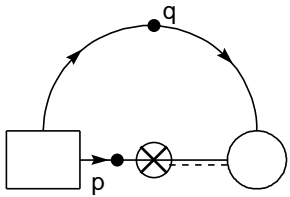}
\end{center}
\caption{\small{Graphical representation of $\DGh(\L_0)$. The box and 
the circle represent the functionals $\bG$ and
$\Pi$ respectively. The top line is the cutoff full propagator of the 
field $\Phi_J$;
the bottom full line represents the field $\Phi_i$ while the double
line is the corresponding source $\g_i$.
The cross denotes the product of the two functionals with the
insertion of the cutoff function $\Kiu(p)$.
Integration over the loop momentum is understood.}}
\end{figure}
 
At this scale $\DG$ is local, or, more precisely,
$\DGi(\L_0)={\cal O}(\frac1{\L_0})$,
so that the irrelevant contributions  disappear in the \UV limit.
From \re{dgb},  $\;\DGb(\L_0)$   
is manifestly relevant, since $\Pi(\L_0)=\Pir(\L_0)$, while 
a direct  calculation shows that 
$\DGh(\L_0)=\DGhr(\L_0)+{\cal O}(\frac1{\L_0})$.
As a matter of fact, from \re{dgh} one notices that
non-local terms may arise from $\bG[\Phi,\g;\L_0]$ and the cutoff
functions. At $\L_0$, $\bG$ is given by either a relevant vertex or 
a sequence of relevant vertices
joint by propagators  with a cutoff function $\Kin(q+P)$, where $P$ is
a combination of external momenta (see \re{bG}).
Since  the integral is damped by these cutoff functions, only the
contributions with a restricted number of propagators survive in the
\UV limit. One can infer from power counting that they are of the 
relevant type. 
A similar argument holds for the possible non-local contributions 
coming from $\Kiu(p)$.
An example of how to compute $\DGh(\L_0)$ at the first loop order will be 
given in appendix A.

\subsection{Solution of $\bom{\DG=0}$ at $\bom{\L=\L_0}$}
Once the locality of $\DG(\L)$ is shown, the solvability of the 
equation $\DG(\L)=0$ at the UV scale is a 
consequence of the consistency condition  ${\cal S}_{\Pi^{(0)}}\,\DG^{(\ell)}
(\L_0)=0$, which in turn follows   from the  $\L$-independence of $\DG$ and 
the solvability of the same equation at $\L=0$ (see ref. \cite{noi}).
The equation $\DG^{(\ell)}(\L_0)=0$ is solved by tuning some of the
relevant couplings in $\Pi^{(\ell)}(\L_0)$. 
In order to see how the procedure works, it is sufficient to consider
the first loop. The generalization to higher loops is straightforward. 
At  one loop and at $\L=\L_0$, $\,\DG$ reads
\beq \nome{deltaglo}
\DG^{(1)}(\L_0)\,=\,\DGr^{(1)}(\L_0)\,={\cal S}_{\Pi^{(0)}}\,
\Pi^{(1)}(\L_0)
\,+\,\DGhr^{(1)}(\L_0)\,.
\eeq
As the first term on the r.h.s. of (\ref{deltaglo}) automatically satisfies 
the consistency condition, this must be the true also for $\DGh^{(1)}$. 
Hence the fine-tuning consists in imposing the relation 
\beq \nome{fintun}
{\cal S}_{\Pi^{(0)}}\, \Pi^{(1)}(\L_0)\,=-\, \DGh^{(1)}(\L_0)\,.
\eeq
Notice that this equation allows to compute the couplings in
$\Pi^{(1)}(\L_0)$ since $\DGh^{(1)}(\L_0)$ depends only on $\Pi^{(0)}$.
At this loop order the line with the crossed circle in Fig.~1 contributes 
only through
a cutoff function $\Kiu$, since the additional full propagator
associated to this line cancels at this order (see \re{dgh}).

The most general functional $\Pi(\L_0)$ contains the relevant monomials
given in \re{mon1} and \re{mon2} and can be split into two contributions
\beq \nome{pigreco1}
\Pi^{(1)}(\L_0)=\Piinv^{(1)}(\L_0)+\Pit^{(1)}
(\L_0)\, ,
\eeq
where $\Piinv$ contains all the independent monomials which are
invariant, \ie ${\cal S}_{\Pi^{(0)}}\, \Piinv^{(1)}=0$. The explicit
form of $\Piinv$ is 
\beeq \nome{piinv}
\Piinv^{(1)}(z_i(\L_0))=\int \mbox{d}^4 x\, \Bigg\{\!\!\!&&\!\!\!\!\!\!\! -
\frac14\,z_1
\, 
{\cal F}^a_{\m\n} \, 
{\cal F}^{a\m\n} + z_2\,z_3 \,\left( \frac1{g\, z_3}\, w^a_\m\, 
{\cal D}^{ab}_\m c^b
-\frac12\, f^{abc}\, v^a\,c^b\,c^c \right)  \\ \nonumber
\!\!\!&&\!\!\!\!\!\!\!\!\!\!\! +\,z_4 \,\bpsi_L\,i\ds{\cal D}\,\psi_L\, +
\, z_5 \,\bpsi_R\,i\ds{\p}\,\psi_R 
\,+\, z_2\,z_3\,\left(\bl\,c\cdot t \, \psi_L\, + 
\,\bpsi_L\,c\cdot t \, \l \right) \Bigg\}\, ,
\eeeq
with ${\cal F}^a_{\mu\nu}=\p_\mu A^a_\nu -\p_\nu A^a_\mu
+ g \, z_3 \, f^{abc} A^b_\mu \,A^c_\nu$ and the covariant derivatives 
given by
${\cal D}^{ab}_\mu c^b =\p_\mu c^a + g \, z_3 \, f^{abc} A^b_\mu \, c^c$ 
and ${\cal D}_\m\,\psi_L=(\p_\m+z_3\, g\, A_\m^a\,\t,a)\,\psi_L$. 
The remaining monomials contribute to $\Pit$ which can be written as
\beeq \nome{pitilde}
\Pit^{(1)} [\Phi,\,\g;\,\s_i(\L_0)]\!\!\!& \equiv&\!\!\! \int\mbox{d}^4x\,
\bigg\{
\s_1\,\L_0^2\,\mbox{Tr}\,(A_\m\,A_\m)\,+\,
\s_2\,\mbox{Tr}\,(\p_\m A_\m\,\p_\n A_\n)\,
%\s_3\,\mbox{Tr}\,(\p_\n A_\m\,A_\m\,A_\n)\,+\,
+\,\s_3\,\mbox{Tr}\,(\p_\m A_\m\,A_\n\,A_\n) \nonumber\\
&+&\!\!\!
\s_4\,\mbox{Tr}\,(A_\m \,A_\m\,A_\n\,A_\n)\,+\,
\s_5\,\mbox{Tr}\,(A_\m \,A_\n\,A_\m\,A_\n)\,+
\s_6\,\mbox{Tr}\,(A_\m \,A_\m)\,\mbox{Tr}\,(A_\n \,A_\n)\,
\nonumber \\
&+&\!\!\!
\s_7\,\mbox{Tr}\,(A_\m \,A_\n)\,\mbox{Tr}\,(A_\m \,A_\n)\,+\,
\s_8\,\mbox{Tr}\,(w_\m\,A_\m\,c)\,+\,
\s_9\, \mbox{Tr} \, (v\,c\,c)\,\nonumber\\
&+&\!\!\! 
\s_{10}\,\bpsi_L\,i\,\ds{A}^a\,\t,a\,\psi_L\,+ \,
\s_{11}\,\bl\,c\cdot t\,\psi_L\,+ \,
\s_{12}\,\bpsi_L\,c\cdot t\,\l\bigg\}\,.
\eeeq
Notice that for $SU(2)$ only nine and for $SU(3)$ only 11 of the monomials 
above are independent.    
Inserting \re{pigreco1} into \re{fintun}, one finds 
\beq \nome{fintun1}
{\cal S}_{\Pi^{(0)}}\, \Pit^{(1)}(\L_0)\,=-\, \DGh^{(1)}(\L_0)\,.
\eeq
which fixes the $\s_i(\L_0)$'s whose finiteness is shown by explicit 
calculation (see Appendix A). 
On the contrary  the couplings $z_i(\L_0)$ 
are not fixed by the fine-tuning, so that one is allowed to set them equal 
to their physical values at $\L=0$, i.e. $z_i(0)=1$. In the standard language
this corresponds to the renormalization prescriptions.

\section{The ABJ anomaly}
The ABJ anomaly represents the breaking of the classical chiral symmetry at 
the quantum level. One is obviously interested in theories where anomalies 
which affect currents coupled to propagating gauge fields
cancel. Nevertheless they have to be computed in order to test the
consistency of the regularization procedure. In this section we
concentrate upon a single left fermion (recall that in our
formulation right fermions are not coupled to the gauge field).
 
As the anomaly is absent at the tree level,  the flow equation
(\ref{chie}) guarantees the $\L$-independence of $\DG$ at 
one loop.
Hence it may be convenient to compute the anomaly at $\L=\L_0$. 
There are  two relevant monomials of $\DG$,  
$\e\m\n\r\s \int \mbox{d}^4x\,
\mbox{Tr}\, \left[c\,\p_\m \,( A_\n\,\p_\r\,A_\s)\right]$ and 
$ \e\m\n\r\s \int \mbox{d}^4x\,
\mbox{Tr}\, \left[c\,\p_\m \,(A_\n\,A_\r\,A_\s)\right]$, 
which are absent 
in $\DGb$ but may be present in $\DGh$. This is due to the  locality of
$\Pi(\L_0)$, which in turn  implies that $\DGb$  is a trivial cocycle
of the cohomology of the BRS operator.
In other words a violation of the ST identity results in the impossibility of
fixing the relevant couplings $\s_i(\L_0)$ in $\Pi^{(1)}(\L_0)$ in such a
way the symmetry is restored, or, equivalently, some of the relevant parameters
in $\DG$ cannot be set to zero. Nevertheless, a consistency condition for them 
still holds (Wess- Zumino condition).

In the following  we compute the fermionic contribution to 
$\DGh$ at one loop order which gives rise to the anomaly. 
Taking  the \UV limit in \re{dgh} and setting  $\L=\L_0$, $\DGh$ 
becomes  
\beeq \nome{primodelta}
\DGh=\DGh^{YM}\!\!\!&+&\!\!
i\int_{p\,q} \frac{\Kin(q)}{q^2}\,\Biggl[\,\ds{q}_{\g \b}\, 
\frac{\de^2 \bG^{(0)}}{\de \psi_\a (-p) \de \bpsi_\b (-q)} \, 
\frac{\de^2 \Pi^{(0)}}{\de \psi_\g (q) \de \bl_\a (p)} \Kiu (p)\Biggr.
\\ \nonumber
& & \;\;\;\;\;\;\;\; + \Biggl.\,\psi\to\bpsi, \; \bl\to\l\;\Biggr]\, .
\eeeq 
In order to compute this functional we need only the tree level
vertices of $\G$, \ie those of $S_{BRS}$, and in particular 
\beeq 
&&\G^{(\bpsi A \psi)}_{\m\,\a\b}(p,\,q,\,-p-q)=ig 
\left(\gm \frac {1-\gc} 2\right)_{\a\b}  \, ,  
\\  \nonumber
&&
\G^{(\bl c \psi)}_{\a \b}(p,\,q,\,-p-q) = \left(\frac{1-\gc}2
\right)_{\a\b} \, , \qquad
\G^{(\bpsi c \l)}_{\a \b}(p,\,q,\,-p-q)= \left(\frac{1+\gc}2
\right)_{\a\b} \,.
\eeeq
The fermionic contribution to the $c$-$A$-$A$ vertex of $\DGh$ is shown in
Fig.~2. Notice that the line with the crossed circle in Fig.~1 contributes 
only through the cutoff function $\Kiu$. 
From \re{primodelta} and \re{bG}  the anomalous part (\ie  the part
giving rise to the antisymmetric tensor  $\e\m\n\r\s$) of this vertex reads 
\beq \nome{an1}
\int_{p\,k}c^a (p)\,A_\n^b (k)\,A_\r^c(r)
\left[\mbox{Tr}\,(\t,a\,\t,b\,\t,c) {\cal A'}^{(cAA)}_{\nu\r}(p,k,r)
+\mbox{Tr}\,(\t,a\,\t,c\,\t,b) {\cal A''}^{(cAA)}_{\nu\r}
(p,k,r)\right]\,,
\eeq
where $r=-p-k$ and 
\beeqn 
&&{\cal A'}^{(cAA)}_{\nu\,\r}(p,\,k,\,r)=
-\,{\cal A''}^{(cAA)}_{\r\,\n}(p,\,k,\,r)\\ \nonumber
&&= \frac{(ig)^2}2
\int_{q}\mbox{Tr}\, \left[\gc (-\ds{q})\,\gn\,(\ds{q}+
\ds{k})\,\gr\,\right] 
  \frac{\Kin (q)}{q^2} \,
\frac{\Kin (q+k)}{(q+k)^2}\,\Kiu (p-q)\,.
\\ \nonumber
\eeeqn

\begin{figure}[htbp]
\epsfysize=4.5cm
\begin{center}
\epsfbox{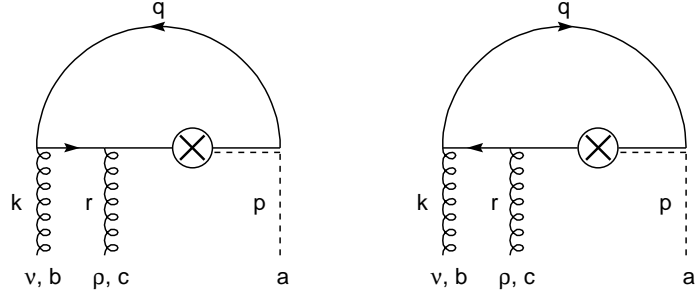}
\end{center}
\caption{\small{Fermionic contribution to the $c$-$A$-$A$ vertex of $\DGh$.
The curly,
dashed and full line denotes the gluon, ghost and fermion field 
respectively; the double lines represent the BRS source
associated to the field depicted by the top line. All momenta are incoming.
}}
\end{figure}

Performing the trace over Dirac matrices, one finds
\beq
{\cal A'}^{(cAA)}_{\nu\,\r}(p,\,k,\,-p-k)=
2\, i\,  g^2 \, \e\a\n\b\r  \,
\int_{q}(q_\a\,q_\b\,+\,q_\a\,k_\b)\, 
\frac{\Kin (q)}{q^2} \,
\frac{\Kin (q+k)}{(q+k)^2}\,\Kiu(p-q)\, .
\eeq
By expanding in the external momenta and taking into 
account the symmetry properties, one obtains 
\beq \nome{terpass}
{\cal A'}^{(cAA)}_{\n\r}\,(p,\,k,\,-p-k)\,=\,\frac{g^2}{16\pi^2}  
\,\e\a\n\b\r \,p_\a\,k_\b\,
\left[\int_0^\infty dx \, \Kin^2(x)\,\frac{\p \Kiu(x)}{\p x} +{\cal 
O}(P^2/\L_0^2)\,\right]\, ,
\eeq
with $x=q^2/\L_0^2$ and  $P$  some combination of the external momenta.
Notice that the result of the integral in \re{terpass} 
is determined only by the values  $\Kin(0)=0$ and $\Kin(\infty)=1$ and
therefore is independent of the choice of the cutoff function.
As a consequence, in the \UV limit, one recovers the usual
contribution to the anomaly,
which is regularization independent. 
In the RG formulation this fact can be understood from the
$\L-$independence of $\DG^{(1)}$, so that the same
result is obtained if one computes the anomaly at 
the physical point $\L=0$. In this case the anomaly comes 
from $\DGb$ and is computed in terms of the physical vertices 
of $\G[\Phi,\g;\L=0]$, which are  regularization independent.

The fermionic contribution to the $c$-$A$-$A$-$A$ vertex of $\DGh$ is shown
in Fig.~3.
According to \re{primodelta} and \re{bG} the  anomalous part of this vertex
reads 
\beeq \nome{an2}
\int_{p\,k\,r}c^a (p)\,A_\n^b (k)\,A_\r^c(r)\,A_\s^d(s)
\!\!\!&&\!\!\!\left[\mbox{Tr}\,(\t,a\,\t,b\,\t,c\,\t,d) {\cal A'}^{(cA
AA)}_{\nu\,\r\,\s}(p,\,k,\,r,\,s)\right.\\ \nonumber
\!\!\!&&\!\!\!+\left.\mbox{Tr}\,(\t,a\,\t,d\,\t,c\,\t,b) {\cal A''}^{(cA
AA)}_{\nu\,\r\,\s}(p,\,k,\,r,\,s)\right]\, ,
\eeeq
where $s=-p-k-r$ and 
\beeq 
{\cal A'}^{(cAAA)}_{\nu\,\r\,\s}(p,\,k,\,r,\,s)\!\!\!&=&\!\!\!
-\,{\cal A''}^{(cAAA)}_{\s\,\r\,\n}(p,\,-(s+p),\,r,\,-(k+p))
\\ \nonumber
\!\!\!&=&\!\!\!\frac{(ig)^3}2
\int_{q}\mbox{Tr}\, \left[\gc (-\ds{q})\,\gn\,(\ds{q}+
\ds{k})\,\gr\,(\ds{q}+\ds{k}+\ds{r})\,\gs \right] \\ \nonumber
\!\!\!&&\qquad \quad \times \frac{\Kin (q)}{q^2} \,
\frac{\Kin (q+k)}{(q+k)^2}\,\frac{\Kin (q+k+r)}{(q+k+r)^2}\,\Kiu (p-q)
\,.
\eeeq

\begin{figure}[htbp]
\epsfysize=4.5cm
\begin{center}
\epsfbox{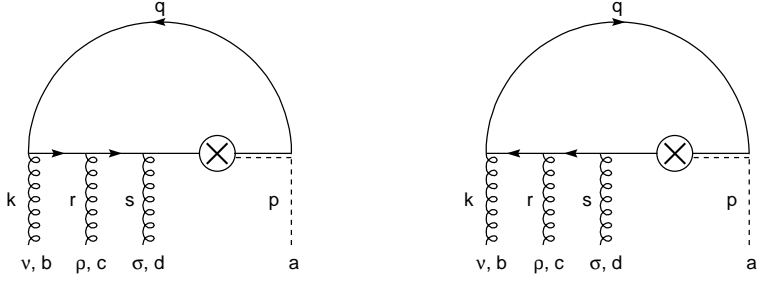}
\end{center}
\caption{\small{Fermionic contribution to the $c$-$A$-$A$-$A$ vertex of 
$\DGh$.
}}
\end{figure}

Performing the trace over the Dirac matrices, one obtains 
\beeq\nome{grafcaaa}
{\cal A'}^{(cAAA)}_{\nu\,\r\,\s}(p,\,k,\,r,\,s)\!\!\!&=&\!\!\!
g^3\,\e\n\r\a\s \,\int_q q^2\, (2\,q_\a\,+\,2\,k_\a\,+r_\a)
\nonumber\\
\!\!\!&\times&\!\!\!\frac{\Kin (q)}{q^2} \,
\frac{\Kin (q+k)}{(q+k)^2}\,\frac{\Kin (q+k+r)}{(q+k+r)^2}\,\Kiu (p-q)\, .
\eeeq
Again, exploiting  symmetry properties and expanding into external 
momenta, one gets
\beeq
&&{\cal A'}^{(cAAA)}_{\nu\,\r\,\s}(p,\,k,\,r,\,s)= \\ \nonumber
&&\frac{i\,g^3}{16\pi^2} \, \e\n\r\a\s\,\int_0^\infty dx 
\left[\frac13\, (2\,k\,+\,r)_\a\,
\frac{\p \Kin^3(x)}{\p x}\,\Kiu(x)\,-\,p_\a\,
\frac{\p \Kiu(x)}{\p x}\,\Kin^3(x)\right]\, ,
\eeeq
where $x=q^2/\L_0^2$ (terms of order ${\cal O}(P^2/\L_0^2)$ are
omitted).
As previously discussed, the integral over $x$ is independent of the
specific cutoff function and in the \UV limit is easily proven to give 
\beq
{\cal A'}^{(cAAA)}_{\nu\,\r\,\s}(p,\,k,\,r,\,s)=
\frac{i}{192\,\pi^2}
\, g^3 \, \e\n\r\a\s\,(2\,k\,+\,r\,+\,3\,p)_\a\,.
\eeq
Combining the two contributions as in (\ref{an2}), 
the $c$-$A$-$A$-$A$ part of the anomaly reads
\beq
\frac{i}{48\,\pi^2}\,g^3\,\e\m\n\r\s \, \int_{p\,k\,r} 
c^a (p)\,A_\n^b (k)\,A_\r^c(r)\,A_\s^d(s) \,
\mbox{Tr}\,(\t,a\,\t,b\,\t,c\,\t,d)\,p_\m\,.
\eeq

There could be in principle a $c$-$A$-$A$-$A$-$A$ vertex in $\DGh$ 
(this monomial is 
also relevant), but it is straightforward to show that the graphs which could 
give rise to such a term vanish because of the $\g$-trace.
Finally, in the coordinate space the anomaly 
has the well-known form \cite{abj}
\beq
{\cal A}=\frac{g^2}{24\,\pi^2}\, \e\m\n\r\s \int \mbox{d}^4x\,
\mbox{Tr}\, \left[c\,\p_\m(A_\n\,\p_\r\,A_\s\,+\,\frac{g}2\,A_\n\,A_\r\,A_\s
)\right]\,.
\eeq 

\section{Conclusions}
In this paper we have considered a CGT within the RG formulation. In
this regularization the cutoff  explicitly breaks  gauge
invariance while  global chiral symmetry is preserved. In the case
of an anomaly  free theory, the Slavnov-Taylor identities are
perturbatively recovered by solving, at the UV scale, a fine-tuning equation 
which fixes all the couplings in the bare action except the five $z_i$'s. 
Then one can exploit such a freedom by fixing these undetermined couplings
at the physical point $\L=0$. In this way the four wave function
normalizations and the three-vector coupling can be set 
at some subtraction point $\mu\ne 0$ to be their physical values $z_i(0)=1$.
In order to recover the complete UV action, in addition to the
fine-tuning procedure one has to perform a perturbative calculation using 
the RG flow. We explicitly computed some of the couplings of this 
UV action. Apart from the $z_i$'s they turn out to be finite. 
This is similar to what happens in dimensional regularization
where the definition of the matrix $\gc$ explicitly breaks the 
chiral symmetry by a term which lives in $\eps$ dimensions 
(the so called evanescent vertex). Once this vertex is inserted
in primitively divergent Feynman diagrams (\ie generating poles in
$\eps$) it produces a finite contribution as $\eps\to 0$. Thus additional
finite counterterms must be introduced in order to restore the 
ST identities. This  fine-tuning involves all possible relevant 
interactions, since the matrix $\g_5$ couples left to right fermions.
On the contrary in our formulation the regularization breaks local 
gauge invariance but preserves global chiral symmetry, so that only 
globally invariant monomials are involved in the 
fine-tuning, which, though  unavoidable,  is thus simplified (the 
situation is even worse with lattice regularization due to
the additional breaking of Lorentz invariance). 

Another advantage of the RG formulation is that it
works in a four dimensional space-time. Hence, as in the case of the 
so-called ``naive'' 
dimensional regularization, the calculation of the $\g$-traces can be easily
performed. In particular, by a simple one-loop calculation, we reproduced
the chiral anomaly result, which turns out to be universal, \ie independent
of the cutoff function choice, as it should
(as well known the naive dimensional regularization cannot
reproduce the chiral anomaly).

We performed a one-loop analysis, but the procedure systematically
generalizes to higher order. This is another advantage with respect to
dimensional regularization, which is difficult to extend to higher
orders due to the ambiguities in handling the matrix $\gc$.

\vspace{1cm}\noindent{\large\bf Acknowledgements}
We have greatly benefited  from discussions with M. Simionato and M. Tonin.

\vspace{1cm}
\section*{Appendix A}
\vskip 0.3 true cm
\noindent
In this appendix we solve the fine-tuning equations at the first loop
order and at the UV scale, $\L=\L_0$, and show that the one loop couplings
$\s_i(\L_0)$ in \re{pitilde} are finite. In this case $\DGb$ contains
the UV couplings of $\Pit^{(1)}(\L_0)$, while the vertices of $\DGh$ 
are given by the product of the tree level vertices of $\bG$ (obtained from
\re{bG}) and those of $\Pi$. 

We first consider the $A$-$c$ vertex of $\DG$. 
From \re{dgb} and \re{pitilde} one can easily see that 
$\bar{\D}_{\G\,\m}^{(Ac)}(p;\L_0)$ is given by 
\beq \nome{deltaac}
\bar{\D}_{\G\,\m}^{(Ac)}(p;\L_0)\,
=-\frac i g \,p_\m\,[\s_1(\L_0)\,\L_0^2+\s_2(\L_0)\,p^2]\,.
\eeq
The graphs contributing to $\hat{\D}_{\G\,\m}^{(Ac)}(p;\L_0)$ are shown 
in Fig.~4.

\begin{figure}[htbp]
\epsfysize=8cm
\begin{center}
\epsfbox{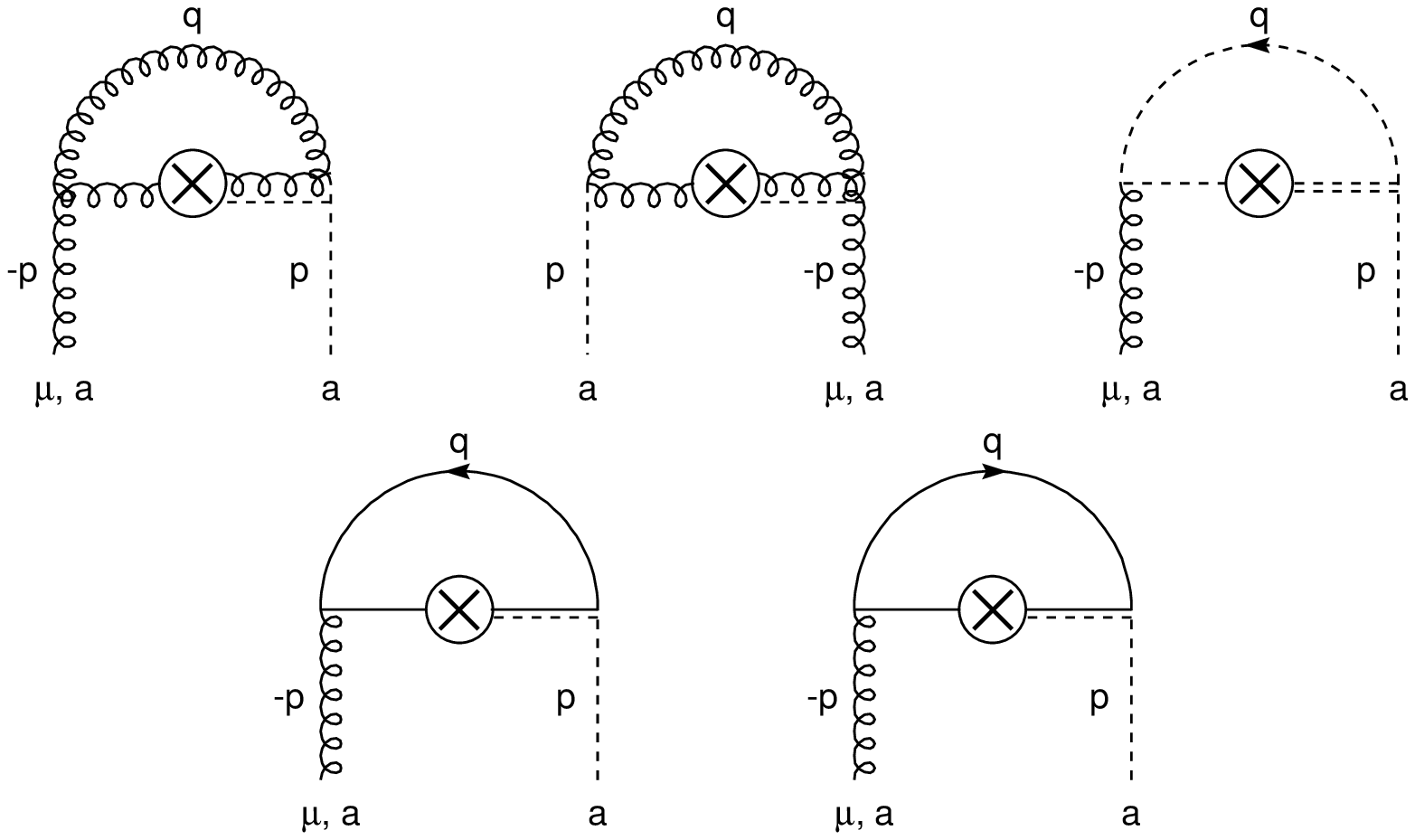}
\end{center}
\caption{\small{First loop contributions to $\hat{\D}_{\G\,\m}^{(Ac)}$.
}}
\end{figure}

\noindent
Using the vertices of $\sbrs$ one can derive the total pure Yang-Mills (YM)
contribution, which  is given by
\beq \nome{grafacym}
\,2\,g \int_q \frac1{q^2}\,\Kiu(p-q)\,\Kin(q)\,(3p-4q)_\m\,;
\eeq
in addition the fermionic (F) contribution is
\beq \nome{grafacfer}
-2\,g\int_q \frac1{q^2} \,\Kiu(p-q)\,\Kin(q)\,\mbox{Tr}\,[\ds{q}\,
\g_\m\, (1-\gc)/2]\,.
\eeq
For  $p\ll\L_0$, one can write
\beq \nome{deltahac}
\hat{\D}_{\G\,\m}^{(Ac)}(p;\L_0)\,= \,p_\m\,[\hat\de_1\,\L_0^2+\hat\de_2\,p^2
\,+\, {\cal O}(p^4/\L_0^2)]\,.
\eeq
where the values of $\hat{\de}_1$ and $\hat{\de}_2$ can be obtained
from \re{grafacym} and \re{grafacfer} once the cutoff function is 
specified. 
We have performed their calculation using three different forms
of $\Kiu$. 

The results are the following (in units of $\frac{ig}{16\,\pi^2}$)
\begin{center}
\begin{tabular}{|l |r | r| r|r |} \hline
  &  &  &  & \\
$\Kiu(p)$ & $\hat{\de}_1^{\mbox{\tiny{(YM)}}}$ &$ 
\hat{\de}_1^{\mbox{\tiny{(F)}}}$& 
$\hat{\de}_2^{\mbox{\tiny{(YM)}}}$&
$\hat{\de}_2^{\mbox{\tiny{(F)}}}$ \\
 & & & & \\  \hline
 & & & & \\
$\Theta (1-p^2/\L_0^2)$ & $2$&1 & -  & -  \\
 & & & & \\
$\frac{\L_0^4}{(p^2+\L_0^2)^2} $&$ \frac23$ &$ \frac53$      
&  $-\frac7{30}$&
$-\frac13$ \\
 & & & & \\
$\exp{\left(-\frac{p^2}{\L_0^2}\right)} $ & 0 &$ \frac32$&$-\frac1{12}$ &
$-\frac13 $  \\

  &  &  &  &   \\
\hline
\end{tabular} 
\end{center}
The fine-tuning equation \re{fintun1}, together with the results 
\re{deltaac} and \re{deltahac}, allows to find the values of
$\s_1(\L_0)$ and $\s_2(\L_0)$ which as a consequence depend on
the cutoff function. The finiteness of the results is due to the presence
in \re{grafacym} and \re{grafacfer} of the two cutoff functions having 
almost non-intersecting supports (\ie $q^2 \gtrsim \L_0^2$, $(q-p)^2 
\lesssim \L_0^2$).
 
In order to check our calculation we have computed these 
relevant couplings at the physical point $\L=0$ using the evolution
equation \re{eveq} and the corresponding cutoff function. In the limit
\UV one has $\s_1(0)=\s_2(0)=0$, as required by the fine-tuning equation
at the physical point $\L=0$ \cite{noi}.

The same analysis can be repeated for the other vertices 
of $\DG$ in order to fix the remaining couplings $\s_i(\L_0)$ in \re{pitilde}. Also these couplings are  finite, thanks to the same argument discussed
above.


\begin{thebibliography}{99}
\bibitem{chiral}
        T. Marinucci and M. Tonin, \nc{31}{381}{76};
        S. Aoyama and M. Tonin, \np{179}{293}{81}; for a review see 
        M. Tonin, Nucl. Phys. B (Proc. Suppl.) 29B,C (1992) 137.
\bibitem{roma} A. Borrelli, L. Maiani, G.C. Rossi, R. Sisto and M. Testa,
        \np{333}{335}{90}; for a review see L. Maiani, Nucl. Phys. B 
        (Proc. Suppl.) 29B,C (1992) 33.    
\bibitem{abj}
        S. Adler and W. A. Bardeen, Phys. Rev. 182 (1969) 1517. 
\bibitem{eps}
        G. t'Hooft and M. Veltman, \np{44}{189}{72}.
        C.G. Bollini and J.J. Giambiagi, \pl{40}{566}{72}; 
        J.F. Ashmore, \lnc{4}{289}{72};
        G.M. Cicuta and E. Montaldi, \lnc{4}{329}{72};
        K.G. Wilson, \pr{7}{2911}{73};
        G. t'Hooft, \np{61}{455}{73};
        E.R. Speer, \jmp{15}{1}{74};
        P. Butera, G.M. Cicuta and E. Montaldi, \nc{19}{513}{74}.
        P. Breitenlohner and D. Maison, \cmp{52}{11, 39 and 55}{77}.
\bibitem{gamma5}
        D.A. Akyeampong and R. Delbourgo, \nc{17}{578}{73}; 18A (1973) 
        94;
        19A (1973) 219;
        T.L. Trueman, \pl{88}{331}{79};
        S. Gottlieb and J.T. Donohue, \pr{20}{3378}{79};
        G. Bonneau, \np{177}{523}{81}.
\bibitem{RSS} G.C. Rossi, R. Sarno and R. Sisto, \np{389}{101}{93}.
\bibitem{F} R. Ferrari, A. Le Yaouanc, L. Oliver and J.-C. Raynal, 
        \pr{52}{3036}{95}.
\bibitem{cohom}
	C. Becchi, A. Rouet and R. Stora, \ap{98}{287}{76};
        G. Costa, J. Julve, T. Marinucci and M. Tonin,
                Il Nuovo Cimento 38A (1977) 373;
        G. Bandelloni, C. Becchi, A. Blasi and R. Collina,
	Ann. Inst. H. Poincar\'e XXVIII (1978) 522; 
                \cmp{72}{239}{80};
	for a review see
	O. Piguet and A. Rouet, \prep{76}{1}{81};
	L. Baulieu, \prep{129}{1}{85}.
\bibitem{Becchi0} C. Becchi, Lectures on the renormalization of gauge 
        theories,
        in {\it Relativity, groups and topology II} (Les Houches 
	1983), Eds.  B.S. DeWitt and R. Stora (Elsevier Science Pub. 1984).
\bibitem{W}
        K.G. Wilson, \prb{4}{3174,3184}{71};
        K.G. Wilson and J.G. Kogut, \prep{12}{75}{74}.
\bibitem{P}
        J. Polchinski, \np{231}{269}{84}.
\bibitem{B}
        C. Becchi, On the construction of renormalized quantum field
        theory using renormalization group techniques,
        in {\it Elementary particles, Field theory and Statistical
        mechanics}, Eds. M. Bonini, G. Marchesini and E. Onofri,
        Parma University 1993.
\bibitem{noi} 
        M. Bonini, M. D'Attanasio and G. Marchesini,
        \np{421}{429}{94}, \np{437}{163}{95}, \pl{346}{87}{95}.
\bibitem{BRS}
        C. Becchi, A. Rouet and R. Stora, \cmp{42}{127}{75},
        \ap{98}{287}{76}.
\bibitem{BDM}
        M. Bonini, M. D'Attanasio and G. Marchesini,
        \np{409}{441}{93}.
\bibitem{Wett}
	C. Wetterich, \pl{301}{90}{93}.
	T. Morris, Int. J. Mod. Phys. A9 (1994) 2411;
	M. Reuter and C. Wetterich, \np{417}{181}{94}.
\bibitem{chi}
        M. Bonini, M. D'Attanasio and G. Marchesini,
        \pl{329}{249}{94}.
\bibitem{Ellw}
	U. Ellwanger, \pl{335}{364}{94}.
\bibitem{MT}
        M. D'Attanasio and T.R. Morris, \pl{378}{213}{96}.

\end{thebibliography}
\end{document}